\documentclass[11pt]{article}

\usepackage[final]{acl}

\usepackage{times}
\usepackage{latexsym}

\usepackage[T1]{fontenc}

\usepackage[utf8]{inputenc}
\usepackage{stfloats}
\usepackage{microtype}

\usepackage{inconsolata}

\usepackage{graphicx}
\usepackage{booktabs}
\usepackage{tabularx}
\usepackage{placeins}

\title{When Should a Virtual Patient Ask? RISP-Guided Proactive Information Seeking in Multi-Agent Patient Simulation}

\author{
\textbf{Jingyao Cai\textsuperscript{1}, Shuaijun Liu\textsuperscript{2}, Abdul Rehman\textsuperscript{1}, Xiaowan Li\textsuperscript{3},} \\
\textbf{Qintian\textsuperscript{4}, Thomas Dolby\textsuperscript{5}, Sue M Green\textsuperscript{1}, Xiaosong Yang\textsuperscript{1}} \\
{\normalfont \textsuperscript{1}Bournemouth University, Bournemouth, UK} \\
{\normalfont \textsuperscript{2}HKUST(GZ), Guangzhou, China\quad \textsuperscript{3}Hiroshima University, Hiroshima, Japan} \\
{\normalfont \textsuperscript{4}Qiongtai Normal University, Haikou, China\quad \textsuperscript{5}i3 Simulations Ltd, UK} \\
{\normalfont\texttt{\{jcai, arehman, sgreen, xyang\}@bournemouth.ac.uk}} \\
{\normalfont\texttt{sliu529@connect.hkust-gz.edu.cn, d233196@hiroshima-u.ac.jp}} \\
{\normalfont\texttt{qintian@mail.qtnu.edu.cn, tom@i3simulations.com}}
}

\begin{document}
\maketitle
\begin{abstract}
Virtual patient (VP) systems support nursing communication training by providing simulated encounters in which learners practise explaining care and responding to patient concerns. Although large language models (LLMs) support flexible dialogue, current approaches give less explicit attention to how patients' information needs develop and guide the content and timing of their questions. We draw on the Risk Information Seeking and Processing (RISP) model, which relates information seeking to perceived risk and the gap between perceived knowledge and desired understanding. Guided by this perspective, we propose the RISP Multi-Agent System (RISP-MAS), a VP system for patient-initiated information seeking. The system links ongoing assessment of the patient's perceptions and information needs to coordinated decisions about whether to seek information, when to ask and which uncertainty to address. Evaluation of nursing student-VP interactions suggested that RISP-MAS exhibited plausible patient information-seeking behaviour, as indicated by expert assessments, while student feedback reflected positive interaction experiences. This work extends patient simulation by connecting a psychological account of information needs to explicit decisions about question initiation. The approach offers nursing students opportunities to practise recognising unresolved patient concerns and tailoring explanations to patients' information needs.
\end{abstract}

\section{Introduction}

Effective communication with patients is a core capability for nursing students and practising nurses, involving listening to concerns, explaining care, and responding to patients' questions \citep{o2014counts}. Nurses' frequent contact with patients creates opportunities for information exchange throughout care \citep{tobiano2015nurses}. These exchanges extend from symptoms and treatment to nursing procedures and everyday concerns, with dialogue and knowledge sharing forming important aspects of patient participation \citep{tobiano2016activities}. Patients may therefore initiate questions about their condition, the purpose of a procedure, or what will happen next. Patient-initiated information seeking is a natural part of nurse-patient communication that learners need opportunities to recognise and respond to.

VP systems provide opportunities to practise these exchanges through simulated nurse-patient encounters. LLMs have expanded the scope for flexible dialogue that reflects clinical cases and patient preferences \citep{cook2025virtual}. This flexibility also supports patient behaviour that adapts to trainee communication as an encounter unfolds \citep{lee2025adaptive}. These developments bring a further requirement for nursing training into focus: representing the information needs that lead patients to initiate questions. Although simulated patients can ask questions, the link between their changing understanding and the content and timing of those questions remains less explicit. Addressing this gap requires modelling how information needs develop during dialogue and guide whether a patient asks, when they ask and which uncertainty they address.

To address this problem, we draw on the RISP model \citep{griffin1999proposed}, which explains risk-related information seeking and processing through factors such as perceived risk, affective response, and information sufficiency. This perspective provides a theoretical basis for modelling patients' information needs as concerns about illness, treatment, or care arise during nurse-patient dialogue. We propose RISP-MAS, a RISP-grounded multi-agent VP system that operationalises these relationships for nursing communication training. Its multi-agent workflow dynamically assesses the patient's perceived understanding and desired information, identifying needs that are relevant to the current interaction. A coordinated seeking decision then determines whether to seek information, selects a specific information goal, and decides when to express it as a patient question.

This work makes three contributions to patient simulation for nursing communication training. First, we adapt RISP to the VP setting to model patient-initiated information seeking from the patient's perceived information needs. Second, we introduce RISP-MAS, which links those needs to decisions about whether to seek information, when to ask, and what to ask about across multi-turn nurse-patient dialogue. Third, we evaluate the system through interactions with nursing students in realistic nursing communication scenarios, combining student feedback with assessments by nursing experts. High scores from both students and nursing experts provide empirical support for the effectiveness of the proposed approach in these scenarios.

\section{Related Work}

LLM-based VPs increasingly combine clinical case representations with control over patient behaviour.
Case-based dialogue simulation can incorporate patient preferences and provide personalised feedback \citep{cook2025virtual}.
PatientSim \citep{kyung2026patientsim} combines clinical profiles with persona dimensions such as personality, language proficiency, recall and confusion.
These dimensions represent variation in doctor-patient interactions.
Multi-agent systems coordinate patient representation, conversational memory and response generation \citep{qu2026large}.
In nursing training, Adaptive-VP \citep{lee2025adaptive} uses assessments of trainee communication to adjust cooperation, resistance and emotional expression.
Across these studies, patient consistency and response adaptation receive particular attention.
However, these mechanisms do not explicitly link a patient's changing understanding to the content and timing of patient-initiated questions.

To represent the thoughts and feelings underlying such behaviour, psychologically grounded simulators make patients' internal states explicit.
PATIENT-$\psi$ \citep{wang2024patient} uses cognitive behavioural therapy models to condition simulated patients' beliefs, thoughts, and emotional responses during therapy role-play.
SimPatient \citep{steenstra2025scaffolding} updates control, self-efficacy, awareness and reward during motivational interviewing training.
It visualises these changes alongside feedback to trainees.
AnnaAgent \citep{wang2025annaagent} models help-seeking clients through evolving emotions and complaints, coupled with memory across counselling sessions.
These states capture therapeutic cognition, emotion and motivation, but do not directly indicate whether a patient feels sufficiently informed about a clinical concern.
The link from changes in emotion or motivation to a specific need for clinical information is therefore not explicitly modelled.

Beyond representing internal states, research on autonomous agents examines how memory and goals guide initiative.
Generative Agents \citep{park2023generative} integrate observations, memory retrieval, reflection, and planning to produce evolving behaviour, including initiating social interactions.
In dialogue, this initiative includes asking for information needed to complete a task.
Ask-before-Plan \citep{zhang2024ask} predicts clarification needs from dialogue and environmental observations within a multi-agent planning framework.
\citet{huang2025teaching} use reinforcement fine-tuning to reward questions that uncover implicit user knowledge and requirements.
These methods frame information gathering around task completion and missing user requirements.
However, these task objectives do not specify whether a patient feels sufficiently informed or how that judgement should determine question content and timing.

\section{Methodology}
\label{sec:methodology}

\subsection{RISP Theoretical Model}
\label{sec:risp-theory}

The RISP model \citep{griffin1999proposed} relates information seeking to individual characteristics, perceived hazard, affect, informational subjective norms, information sufficiency, perceived information gathering capacity (PIGC) and channel beliefs.

Perceived hazard refers to an individual's perception of potential threats and their consequences, while affect captures associated emotional responses, such as worry or anxiety. Informational subjective norms reflect perceived expectations from others to acquire relevant information.

Information sufficiency concerns whether perceived knowledge meets the level of understanding an individual feels they need \citep{kahlor2003studying}. RISP-MAS assesses this gap qualitatively for specific concerns, such as symptom cause or treatment.

PIGC refers to the perceived ability to obtain and understand needed information, whereas channel beliefs concern the perceived reliability and usefulness of information sources. In nurse-patient interactions, these beliefs can relate to the patient's confidence in asking questions and trust in the nurse as an information source.

For example, a patient concerned about a potential complication may feel anxious and recognise a need to understand its causes, while perceived ability to ask questions and trust in the nurse may influence whether the patient seeks clarification.

\begin{figure}[t]
  \centering
  \includegraphics[width=\columnwidth]{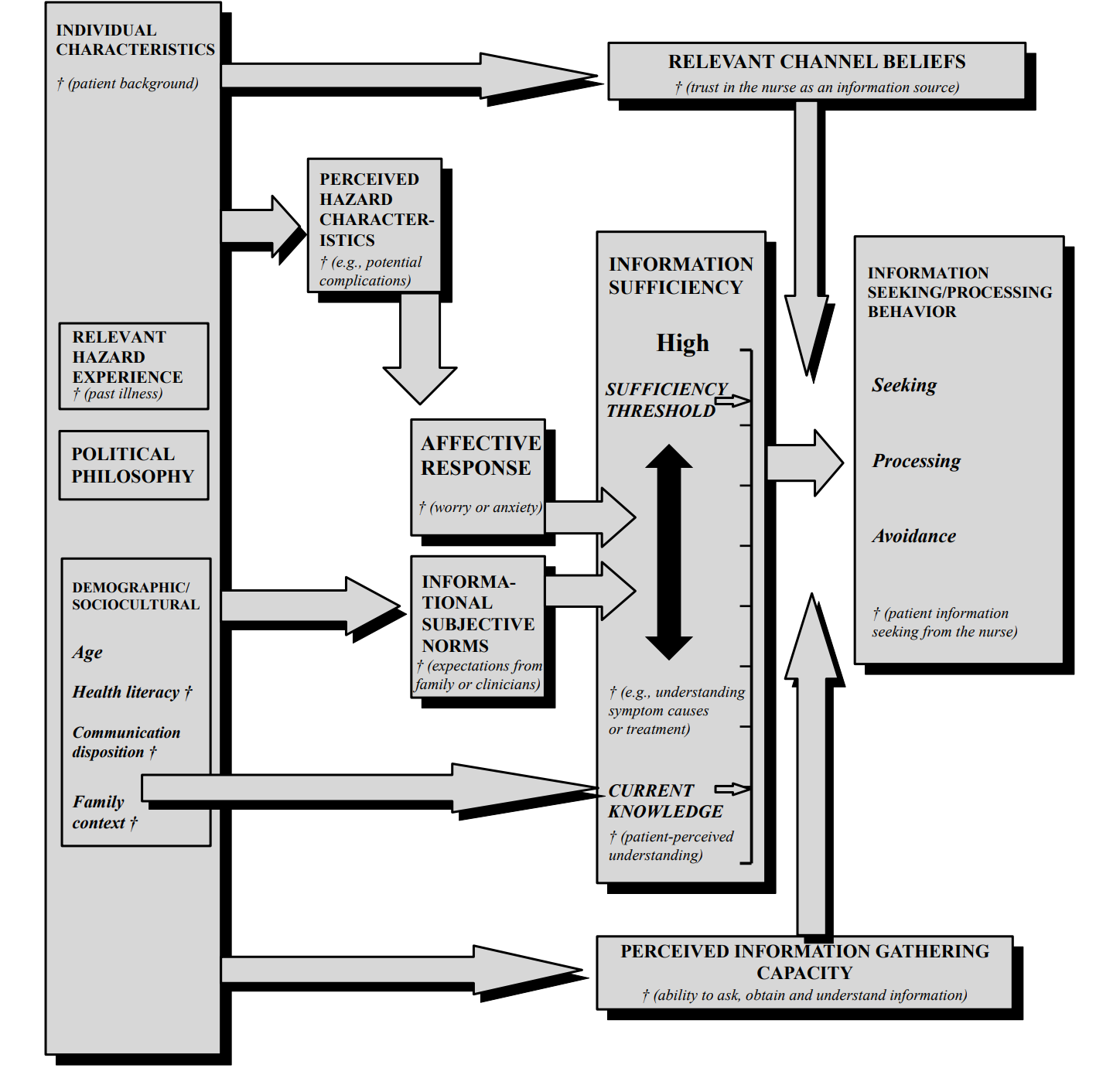}
  \caption{The RISP model, adapted from \citet{griffin1999proposed} for VP simulation. Annotations marked with $\dagger$ provide patient-specific examples and contextual interpretations for nurse-patient interactions.}
  \label{fig:risp-theoretical-model}
\end{figure}

These factors provide the theoretical basis for assessing patient information needs and guiding information-seeking decisions in RISP-MAS. Figure~\ref{fig:risp-theoretical-model} presents the theoretical factors underlying these assessments.

\subsection{RISP-MAS}
\label{sec:risp-architecture}

\begin{figure*}[t]
  \centering
  \includegraphics[width=\textwidth]{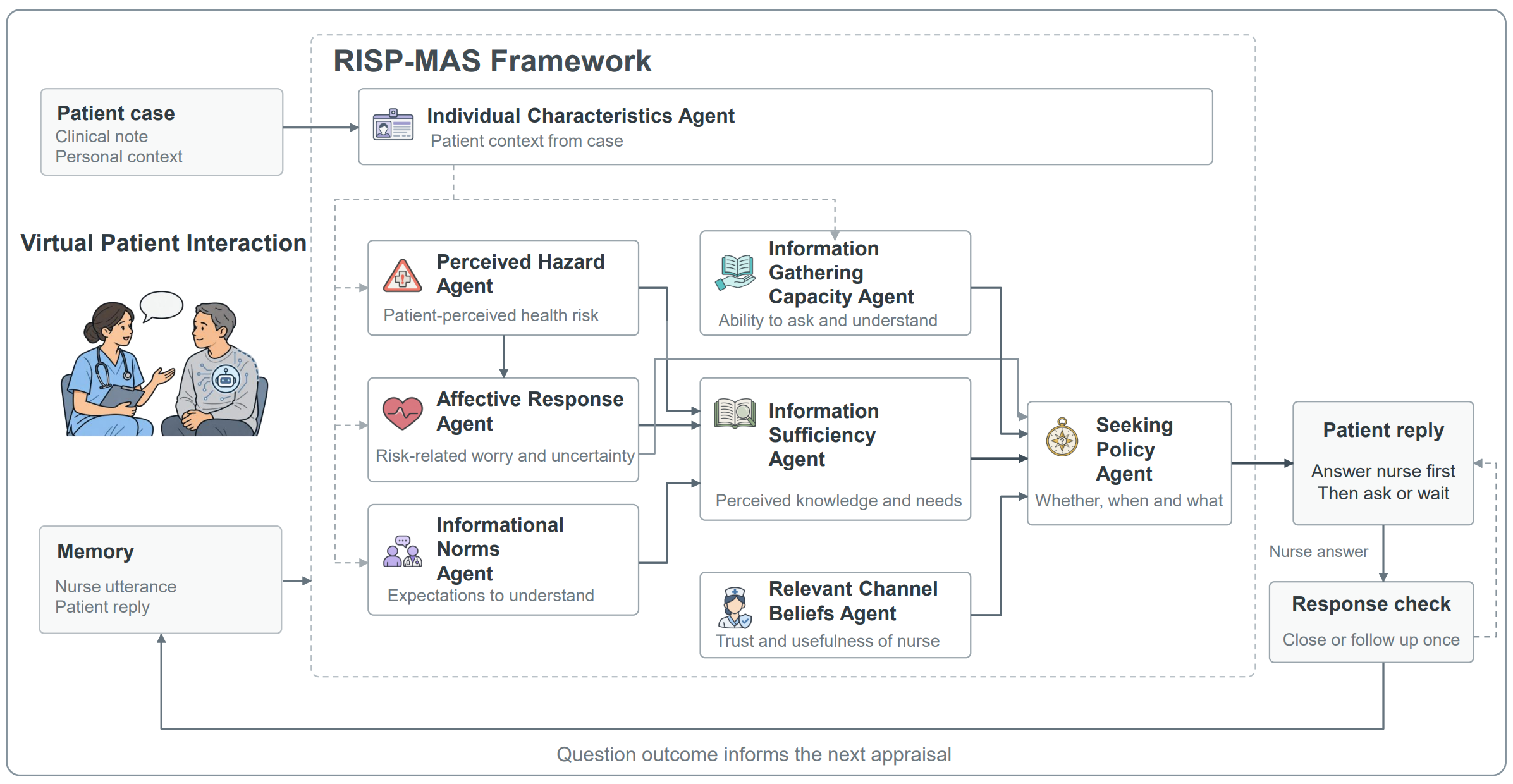}
  \caption{RISP-MAS architecture for patient question initiation.}
  \label{fig:risp-mas-architecture}
\end{figure*}

Figure~\ref{fig:risp-mas-architecture} illustrates how eight agents coordinate RISP appraisals and patient question initiation.
The Individual Characteristics Agent derives patient characteristics from the case.
4 agents assess perceived hazard, informational norms, PIGC and channel beliefs.
The Affective Response Agent assesses risk-related emotions, the Information Sufficiency Agent identifies information needs, and the Seeking Policy Agent decides whether, when and what to ask.

For exchange $e_t=(u_t,y_t)$, where $u_t$ is the nurse utterance and $y_t$ the patient reply, the RISP representation is
\begin{equation}
  S_t=(I,R_t,A_t,N_t,G_t,B_t,\mathcal{N}_t,D_t),
  \label{eq:risp-state}
\end{equation}
Here, $I$ represents case-derived patient characteristics; $R_t$, $A_t$ and $N_t$ represent perceived hazard, risk-related affect and informational norms; and $G_t$ and $B_t$ represent PIGC and channel beliefs.
$\mathcal{N}_t$ contains identified information needs, while $D_t$ records the seeking decision and its rationale.
Initially, $\mathcal{N}_0=\emptyset$ indicates that no needs have yet been registered, rather than an absence of pre-existing patient concerns.

Appraisals are updated following nurse-patient exchanges and inform subsequent patient replies.
While a patient question is being expressed or answered, RISP appraisal is paused.
A response check determines whether the nurse adequately addresses the question.
An adequate response closes the question; otherwise, one follow-up on the same goal is permitted before further questioning about it stops.
RISP appraisal then resumes on the following turn, incorporating the exchanges and question outcome.

\subsubsection{Individual Characteristics Agent}
\label{sec:individual-characteristics}

The Individual Characteristics Agent derives patient characteristics $I$ from the case information available to the simulated patient.
These include experiences with illness or medical treatment, familiarity with health information, communication disposition, and family or social circumstances described in the case.
These characteristics inform subsequent appraisals of perceived health risks, related emotional responses, perceived expectations to seek health information, and the patient's perceived ability to obtain and understand such information.

\subsubsection{Perceived Hazard Agent}
\label{sec:perceived-hazard}

The Perceived Hazard Agent assesses how the patient perceives potential health threats associated with their condition or treatment, including perceived seriousness, susceptibility and personal control.
It uses the previous appraisal of the patient's perceived health risk, case-derived patient characteristics, active information needs and relevant current or earlier exchanges.
The output contains ordinal ratings for these perceptions and a summary of the patient's perceived health risk, with cited supporting evidence.
Perceptions remain unchanged when the available evidence does not support an update.
The resulting $R_t$ informs the Affective Response Agent's assessment of the patient's emotional reactions, the Information Sufficiency Agent's identification of information needs and the Seeking Policy Agent's decision on question initiation.

\subsubsection{Affective Response Agent}
\label{sec:affective-response}

The Affective Response Agent assesses risk-related worry, anger and uncertainty using the hazard appraisal, previous affect, patient characteristics and relevant exchanges:
\begin{equation}
  A_t=f_{\mathrm{aff}}(A_{t-1},R_t,I,U_{t-1},e_t),
  \label{eq:risp-affect}
\end{equation}
Here, $U_{t-1}$ contains earlier exchanges and patient question outcomes available for the appraisal, and $f_{\mathrm{aff}}$ denotes the structured LLM assessment.
The agent returns the current levels of worry, anger and uncertainty, with evidence supporting any change.
Worry concerns personally relevant harm, while anger concerns the source or handling of the risk.
Uncertainty concerns the risk's nature, development, consequences or controllability.
$A_t$ informs information sufficiency and the decision to seek information.
Worry can persist after the patient has gained sufficient understanding of a particular topic.

\subsubsection{Informational Norms Agent}
\label{sec:informational-norms}

The Informational Norms Agent identifies expectations from important others that make the patient feel they should understand a risk-related topic.
It uses previous norms, patient characteristics and explicit cues in current or earlier exchanges, including evidence from patient questions.
The output gives an ordinal norm level and identifies the sources of the expectation, such as family members or clinicians.
Norms remain unchanged when the evidence does not support an update.
$N_t$ informs information sufficiency through explicit expectations about seeking understanding.
General advice from family members or clinicians alone does not establish an informational subjective norm unless it indicates that the patient feels expected to seek or understand health information.

\subsubsection{Information Sufficiency Agent}
\label{sec:information-sufficiency}

The Information Sufficiency Agent identifies what the patient currently feels they need to know.
It assesses perceived understanding, desired information and the current relevance of a concern before judging the sufficiency threshold and remaining information insufficiency.
The resulting needs are represented as
\begin{equation}
  \begin{array}{rl}
    \widehat{\mathcal{N}}_t={}&f_{\mathrm{suf}}(\mathcal{N}_{t-1},R_t,A_t,N_t,\\
                            &P_{\mathrm{need}},H_t),\\
    \mathcal{N}_t={}&\mathrm{Update}(\mathcal{N}_{t-1},\widehat{\mathcal{N}}_t,H_t),
  \end{array}
  \label{eq:risp-sufficiency}
\end{equation}
Here, $\widehat{\mathcal{N}}_t$ contains appraised needs and their cited evidence, while $f_{\mathrm{suf}}$ denotes the structured LLM assessment.
$\mathrm{Update}$ incorporates these appraisals into $\mathcal{N}_t$ and determines which needs are active or dormant.
$P_{\mathrm{need}}$ contains the psychological-status description available to the patient, and $H_t$ contains relevant conversation history and question outcomes.
A concern stated in the case becomes active when an exchange establishes its current relevance and the information goal remains unanswered.
An information goal associated with a previously closed question is excluded.
Each need addresses one uncertainty and records perceived understanding, knowledge, the desired sufficiency threshold, information insufficiency and salience.
The appraisal normally introduces at most one new active need, unless the patient explicitly states several independent current concerns.
The LLM assesses insufficiency using ordinal categories, without subtracting knowledge ranks.
A need remains active when the concern is currently salient and perceived understanding is insufficient.
The need representation retains at most three active needs, prioritising higher salience and then greater insufficiency.
Earlier needs that are omitted from the appraisal remain recorded as dormant.
After a question ends, its associated need becomes dormant in the next appraisal.
For a resolved question, the nurse's response informs perceived knowledge.
For an abandoned question, unanswered content remains unchanged.
A separate treatment or next-step concern can remain active when its own uncertainty is still salient.
PIGC and channel beliefs inform the subsequent seeking decision rather than this appraisal.

\subsubsection{Perceived Information Gathering Capacity Agent}
\label{sec:pigc}

The PIGC Agent assesses whether the patient feels able to ask for, obtain and understand information in the current interaction.
It uses previous capacity, patient characteristics, health literacy and relevant current or earlier exchanges.
The output contains an ordinal capacity level and supporting evidence of barriers or facilitators.
Capacity remains unchanged when the evidence does not support an update.
$G_t$ informs whether the patient expresses an identified need.
This appraisal represents the patient's perceived ability to gather information from the nurse.

\subsubsection{Relevant Channel Beliefs Agent}
\label{sec:channel-beliefs}

The Relevant Channel Beliefs Agent assesses the perceived trustworthiness and usefulness of the nurse as an information source.
It uses previous channel beliefs and relevant current or earlier exchanges to produce ordinal appraisals for the current nurse.
Channel beliefs are updated only when conversational evidence indicates a change in the patient's perceived trustworthiness or usefulness of the nurse; otherwise, the previous beliefs are retained.
$B_t$ informs whether the patient seeks information from that nurse.

\subsubsection{Seeking Policy Agent}
\label{sec:seeking-policy}

The Seeking Policy Agent decides whether to express one active information need or continue waiting.
It considers need salience, insufficiency, perceived understanding and the sufficiency threshold, alongside risk, affect, PIGC, channel beliefs and conversational context.
The context includes recent exchanges, nurse commitments, previous question goals and outcomes, and how long the need has remained active.
The agent also considers changes in the patient's perceived knowledge and recorded understanding for each active information need. Knowledge is represented by an ordinal category, with an increase from the previous appraisal indicating meaningful progress. Changes in the recorded understanding without an increase in knowledge category indicate minimal progress.
An unchanged record or a single observation is classified as showing no progress.
A persistent, salient need with little knowledge progress can support asking, without a fixed turn threshold.
The agent distinguishes evidence of an imminent explanation from routine assessment when judging whether the need will be addressed naturally.
It returns a seek-or-wait decision, the chosen need and judgments about appropriate timing and justification.
A question is initiated when the decision $d_t$ satisfies
\begin{equation}
  \begin{array}{rl}
    c_t={}&(d_t=\mathrm{seek})\ \land\ \tau_t\ \land\ \xi_t\\
           &\land\ (i_t\in\mathcal{I}_t)\ \land\ \epsilon_t\ \land\ \neg\alpha_t,
  \end{array}
  \label{eq:seeking-gate}
\end{equation}
Here, $i_t$ identifies the chosen need and $\mathcal{I}_t$ contains the identifiers of active needs.
$\tau_t$ denotes appropriate timing, while $\xi_t$ indicates whether expressing the selected information need is warranted by its unresolved status and conversational context.
$\epsilon_t$ denotes the availability of valid supporting evidence references.
$\alpha_t$ indicates whether the information goal matches a previously closed question by need identifier or goal topic.

Timing is assessed from the current exchange.
Urgent, administrative or unrelated exchanges, and an explicit commitment to explain immediately, are treated as unsuitable moments for initiating the question.
The chosen need supplies the information goal.
On the next dialogue turn, the patient answers the nurse first and then expresses that goal as one question.
The Seeking Policy Agent chooses the goal and its timing, while dialogue generation produces the spoken question.

\section{Experiment Setup}
\label{sec:experiment-setup}

\subsection{Patient Profiles}
\label{sec:patient-profiles}

We used 30 synthetic patient profiles prepared by nursing experts from common scenarios in nursing textbooks.
Each profile contained only the information specified in the corresponding scenario, without additional background details.
Clinical information described the patient's condition and care context, while personal information described their relevant background and circumstances.
These profiles provided the case inputs for the simulated interactions.

\subsection{Evaluation Criteria}
\label{sec:evaluation-criteria}

The six criteria were Question Appropriateness (QA), Question Timing Appropriateness (QTA), Information Need Alignment (INA), Proactive Engagement (PE), Behaviour Following Question Resolution (BFQR) and Overall Quality (OQ). OQ is rated independently as a holistic assessment of patient information-seeking behaviour, rather than calculated as the average of the other five criterion scores.
Each criterion was rated on a five-point Likert scale (1-5), with higher scores indicating better quality.
Full definitions and scoring descriptions are provided in Appendix~\ref{sec:expert-evaluation-criteria}.
Rating summaries report the mean, standard deviation (SD) and 95\% confidence interval (CI).
Expert-judge alignment uses Spearman's rank correlation coefficient ($\rho$) and mean absolute error (MAE).

\subsection{Evaluation Protocol}
\label{sec:evaluation-protocol}

We recruited 40 undergraduate nursing students.
Each student completed one interaction with a randomly assigned patient profile, yielding 40 completed student-patient interactions with RISP-MAS.
Each completed interaction was treated as one case for evaluation.

\paragraph{Expert evaluation.}
Eight nursing experts independently evaluated the 40 completed student-patient interactions, one from each of the 40 nursing students.
The interactions were randomly allocated to the experts, with three experts assessing each interaction and each expert assessing 15 interactions.
This allocation yielded 120 assessments, each covering all six criteria.
The evaluation procedure is described in Appendix~\ref{sec:expert-procedure}.

\paragraph{LLM-as-Judge evaluation.}
We used four LLM judges: GPT-6.1 Sol, Claude Opus 5.5, Gemini 3.1 Pro and DeepSeek-V4-Pro.
Model selection considered both model capability and evaluation cost.
The selected models were considered broadly comparable in these two respects.
The judges included one model from the generating model's vendor and three from other vendors to reduce reliance on assessments from a single vendor.
Each judge independently scored the same 40 completed student-patient interactions once using the six criteria and five-point Likert scale applied by the nursing experts.
For each interaction, all judges received the same patient profile and full nurse-patient dialogue.
The generating model's identity, expert scores and other judges' scores were excluded from their inputs.
This comparison examined how automated ratings varied with the choice of judge.


\paragraph{Student feedback.}
After the interaction, students were invited to complete an online questionnaire about their interaction and learning experience.
The analysis included 40 student questionnaires.
The three items assessed patient initiative, perceived learning, and satisfaction and willingness to reuse the platform.
Items were scored on a five-point Likert scale, and responses of 4 or 5 were counted as positive.
The student procedure and full questionnaire are provided in Appendix~\ref{sec:student-procedure} and Appendix~\ref{sec:student-questionnaire}, respectively.



\paragraph{Statistical analysis.}
Expert and LLM ratings used case-level percentile-bootstrap CIs; each resampled expert case retained all three ratings.
Questionnaire CIs were estimated using percentile bootstrap resampling at the student level.
The LLM and questionnaire analyses each used 10,000 bootstrap iterations.
For expert-judge alignment, we matched the 40 cases by their original case identifiers and averaged the three expert ratings per case and criterion.
Spearman's $\rho$ and MAE compared each judge with this expert reference.

\subsection{Implementation Details}
\label{sec:implementation-details}

 RISP-MAS was implemented in LangGraph \citep{langchain2024langgraph} to coordinate the appraisal agents and patient dialogue.
 The system used OpenAI GPT-5.6 Luna \citep{openai_gpt56luna} with the reasoning effort set to medium.
 Model selection considered response latency and the cost of repeated agent calls during interactive simulation.
 GPT-6.1 Sol \citep{openai_gpt61sol}, Claude Opus 5.5 \citep{anthropic2026opus55} and Gemini 3.1 Pro \citep{google2026gemini31pro} used the medium reasoning setting; DeepSeek-V4-Pro \citep{xu2026deepseek} used high.
The agent and judge prompt templates are provided in Appendix~\ref{sec:prompts}.

\begin{table}
  \centering
  \small
  \begin{tabular}{lcc}
    \hline
    \textbf{Criterion} &
    \textbf{Mean $\pm$ SD} &
    \textbf{95\% CI} \\
    \hline
    QA  & 4.450 $\pm$ 0.548 & [4.325, 4.575] \\
    QTA & 4.067 $\pm$ 0.604 & [3.942, 4.183] \\
    INA & 4.567 $\pm$ 0.604 & [4.408, 4.708] \\
    PE  & 4.275 $\pm$ 0.635 & [4.108, 4.433] \\
    BFQR & 4.108 $\pm$ 0.632 & [3.983, 4.242] \\
    OQ  & 4.042 $\pm$ 0.525 & [3.933, 4.150] \\
    \hline
  \end{tabular}
  \caption{%
    Nursing expert ratings of patient information-seeking behaviour across 40 cases, with three experts per case (120 ratings). Entries report means, SDs and 95\% CIs on the five-point Likert scale.
  }
  \label{tab:expert-results}
\end{table}

\section{Experiment Results}
\label{sec:experiment-results}



 \subsection{Expert Evaluation}
\label{sec:expert-results}

\begin{table*}[!b]
  \centering
  \small
  \setlength{\tabcolsep}{3pt}
  \resizebox{\textwidth}{!}{%
  \begin{tabular}{lcccccccc}
    \hline
    \textbf{Criterion}
    & \multicolumn{2}{c}{\textbf{GPT-6.1 Sol}}
    & \multicolumn{2}{c}{\textbf{DeepSeek-V4-Pro}}
    & \multicolumn{2}{c}{\textbf{Claude Opus 5.5}}
    & \multicolumn{2}{c}{\textbf{Gemini 3.1 Pro}} \\
    \cline{2-3}\cline{4-5}\cline{6-7}\cline{8-9}
    & \textbf{Mean $\pm$ SD} & \textbf{95\% CI}
    & \textbf{Mean $\pm$ SD} & \textbf{95\% CI}
    & \textbf{Mean $\pm$ SD} & \textbf{95\% CI}
    & \textbf{Mean $\pm$ SD} & \textbf{95\% CI} \\
    \hline

    QA
    & 4.875 $\pm$ 0.463 & [4.725, 5.000]
    & 4.600 $\pm$ 1.081 & [4.250, 4.900]
    & 3.875 $\pm$ 0.686 & [3.650, 4.075]
    & 4.600 $\pm$ 1.105 & [4.225, 4.900] \\

    QTA
    & 4.775 $\pm$ 0.530 & [4.600, 4.925]
    & 4.400 $\pm$ 1.172 & [4.025, 4.725]
    & 3.500 $\pm$ 0.641 & [3.300, 3.700]
    & 4.600 $\pm$ 1.105 & [4.225, 4.900] \\

    INA
    & 4.700 $\pm$ 0.853 & [4.425, 4.925]
    & 4.500 $\pm$ 1.281 & [4.075, 4.850]
    & 3.575 $\pm$ 0.984 & [3.250, 3.850]
    & 4.625 $\pm$ 1.102 & [4.250, 4.925] \\

    PE
    & 4.325 $\pm$ 1.207 & [3.925, 4.675]
    & 4.375 $\pm$ 1.334 & [3.925, 4.750]
    & 2.900 $\pm$ 0.871 & [2.625, 3.150]
    & 4.450 $\pm$ 1.260 & [4.050, 4.800] \\

    BFQR
    & 4.600 $\pm$ 0.709 & [4.375, 4.800]
    & 4.500 $\pm$ 1.132 & [4.125, 4.801]
    & 3.900 $\pm$ 0.591 & [3.725, 4.075]
    & 4.400 $\pm$ 1.194 & [4.000, 4.750] \\

    OQ
    & 4.325 $\pm$ 0.859 & [4.050, 4.575]
    & 4.350 $\pm$ 1.292 & [3.925, 4.725]
    & 3.175 $\pm$ 0.594 & [3.000, 3.350]
    & 4.425 $\pm$ 1.174 & [4.050, 4.750] \\

    \hline
  \end{tabular}%
  }
  \caption{%
    Patient information-seeking quality assessed by 4 LLM judges using the six expert criteria. Entries report means, SDs and 95\% CIs on the five-point Likert scale for the same 40 cases. Each judge independently supplied one rating per case and criterion.
  }
  \label{tab:llm-judge-results}
\end{table*}

RISP-MAS received favourable nursing expert ratings across all six criteria (Table~\ref{tab:expert-results}).
All criterion means exceeded 4.0, placing them above the \textit{Moderate} scale anchor (3).
The mean OQ score was 4.042 (95\% CI [3.933, 4.150]), reflecting a positive overall assessment of patient information-seeking behaviour.

INA received the highest mean score of 4.567, followed by QA at 4.450.
These ratings indicated that generated questions generally reflected plausible patient uncertainties and remained relevant to the clinical exchange.
Favourable PE and QTA ratings further supported active information seeking at appropriate points in the dialogue.
BFQR also received favourable ratings, indicating that patients generally adjusted or stopped repeated questions after receiving adequate information.
Together, these findings supported contextually appropriate question initiation and resolution in the evaluated nursing scenarios.

\begin{figure*}[t]
  \centering
  \includegraphics[width=\textwidth]{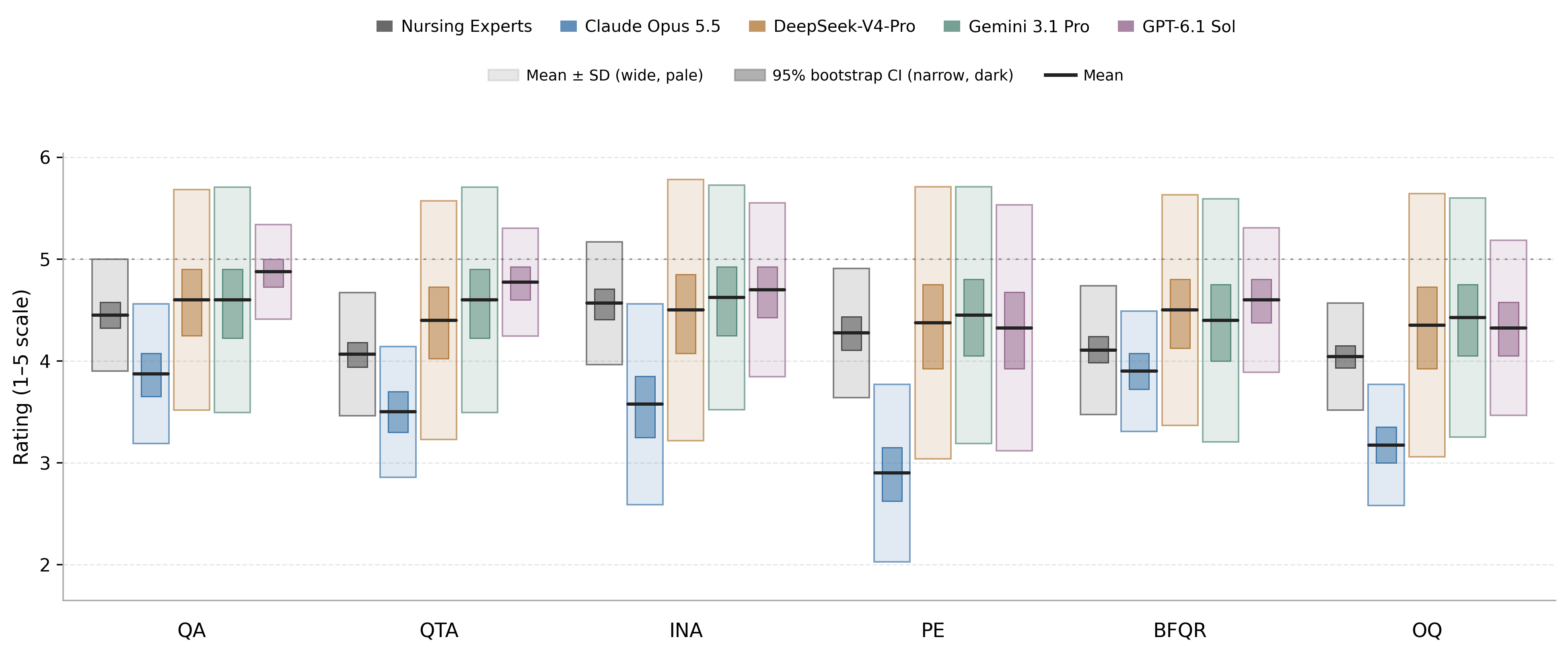}
  \caption{%
    Expert and LLM ratings across six criteria.
    Black lines mark means, wide pale bands show mean $\pm$ SD,
    and narrow dark bands show 95\% percentile-bootstrap CIs for the mean.
    The 40 cases contributed 120 expert ratings (three per case)
    and 40 ratings per LLM judge (one per case).
    Ratings are bounded by 1 and 5; mean $\pm$ SD bands can extend beyond these limits.
  }
  \label{fig:expert-llm-rating-profiles}
\end{figure*}

\begin{table*}[!t]
  \centering
  \small
  \setlength{\tabcolsep}{4pt}
  \resizebox{\textwidth}{!}{%
  \begin{tabular}{lcccccccc}
    \hline
    \textbf{Criterion}
    & \multicolumn{2}{c}{\textbf{GPT-6.1 Sol}}
    & \multicolumn{2}{c}{\textbf{DeepSeek-V4-Pro}}
    & \multicolumn{2}{c}{\textbf{Claude Opus 5.5}}
    & \multicolumn{2}{c}{\textbf{Gemini 3.1 Pro}} \\
    \cline{2-3}\cline{4-5}\cline{6-7}\cline{8-9}
    & \textbf{Spearman's $\rho$} & \textbf{MAE}
    & \textbf{Spearman's $\rho$} & \textbf{MAE}
    & \textbf{Spearman's $\rho$} & \textbf{MAE}
    & \textbf{Spearman's $\rho$} & \textbf{MAE} \\
    \hline

    QA
    & 0.422 & 0.492
    & 0.587 & 0.600
    & 0.706 & 0.575
    & 0.584 & 0.600 \\

    QTA
    & 0.542 & 0.742
    & 0.497 & 0.833
    & 0.796 & 0.567
    & 0.493 & 0.950 \\

    INA
    & 0.596 & 0.400
    & 0.623 & 0.533
    & 0.713 & 0.992
    & 0.596 & 0.475 \\

    PE
    & 0.708 & 0.600
    & 0.696 & 0.683
    & 0.740 & 1.375
    & 0.660 & 0.725 \\

    BFQR
    & 0.537 & 0.708
    & 0.364 & 0.942
    & 0.627 & 0.408
    & 0.561 & 0.875 \\

    OQ
    & 0.752 & 0.517
    & 0.633 & 0.892
    & 0.793 & 0.867
    & 0.625 & 0.867 \\

    \hline
  \end{tabular}%
  }
  \caption{%
    Alignment between 4 LLM judges and mean expert ratings across six criteria (40 matched cases). Spearman's $\rho$ measures ranking similarity, while MAE measures numerical rating differences. 
  }
  \label{tab:expert-llm-alignment}
\end{table*}

\begin{table*}[!t]
  \centering
  \small
  \setlength{\tabcolsep}{5pt}
  \begin{tabular}{lp{0.42\textwidth}ccc}
    \hline
    \textbf{Item} &
    \textbf{Evaluation focus} &
    \textbf{Mean $\pm$ SD} &
    \textbf{95\% CI} &
    \textbf{Positive (\%)} \\
    \hline

    1 &
    Patient Initiative&
    4.100 $\pm$ 0.810 &
    [3.850, 4.325] &
    87.5 \\[0.6ex]

    2 &
    Perceived Learning  &
    4.150 $\pm$ 0.949 &
    [3.850, 4.425] &
    77.5 \\[0.6ex]

    3 &
    Satisfaction and Reuse&
    4.450 $\pm$ 0.639 &
    [4.250, 4.625] &
    97.5 \\

    \hline
  \end{tabular}
  \caption{%
    Nursing student feedback after VP interactions, with one questionnaire per student ($n=40$). Entries report means, SDs, 95\% CIs and positive response percentages (ratings of 4 or 5 on a five-point scale).
  }
  \label{tab:student-results}
\end{table*}

\subsection{LLM-as-Judge Evaluation}
\label{sec:llm-judge-results}

GPT-6.1 Sol, DeepSeek-V4-Pro and Gemini 3.1 Pro assigned mean scores above 4 on all six criteria, indicating favourable assessments of RISP-MAS (Table~\ref{tab:llm-judge-results}).
These three judges also showed favourable mean rating profiles in Figure~\ref{fig:expert-llm-rating-profiles}.
Claude Opus 5.5 gave lower mean ratings, but its OQ mean of 3.175 remained above the \textit{Moderate} scale anchor (3).
Its lowest mean score was for PE (2.900), indicating a less favourable assessment of proactive engagement.
Figure~\ref{fig:expert-llm-rating-profiles} showed favourable mean rating profiles for GPT-6.1 Sol, DeepSeek-V4-Pro and Gemini 3.1 Pro, with lower mean ratings from Claude Opus 5.5.

Because RISP-MAS used GPT-5.6 Luna, we included Claude Opus 5.5, DeepSeek-V4-Pro and Gemini 3.1 Pro alongside GPT-6.1 Sol.
This cross-evaluation reduced reliance on a possible preference for outputs from the same vendor.
All 4 models gave favourable overall ratings, supporting the positive assessment of RISP-MAS's patient information-seeking behaviour.

\subsection{Expert-LLM Rating Alignment}
\label{sec:expert-llm-alignment}

All 4 LLM judges showed positive rank correlations with nursing expert ratings across the six criteria (Table~\ref{tab:expert-llm-alignment}).

For OQ, Spearman's $\rho$ ranged from 0.625 to 0.793, indicating that the judges generally ranked cases similarly to the experts.
OQ MAE ranged from 0.517 to 0.892, below one rating point for all 4 judges on the five-point scale.
These findings supported correspondence in both OQ rankings and scores.
This agreement complemented the favourable expert and LLM assessments, supporting the positive overall evaluation of RISP-MAS.

\subsection{Student Evaluation}
\label{sec:student-results}

Nursing students gave favourable ratings across the three feedback items (Table~\ref{tab:student-results}).
Patient Initiative received a mean score of 4.100, with 87.5\% positive responses.
Most students therefore reported noticing the simulated patient initiating questions or concerns.

Perceived Learning received a mean score of 4.150, with 77.5\% positive responses.
Satisfaction and Reuse received the highest mean score of 4.450, with 97.5\% positive responses.
These responses indicated perceived learning value and satisfaction with the interaction.

\section{Conclusion}

We introduced RISP-MAS, a multi-agent VP system grounded in the RISP model for patient-initiated information seeking.
 Its agents link changing information needs to decisions about whether to seek information, when to ask and what to ask about.
 Expert ratings supported plausible and contextually appropriate patient information-seeking behaviour in the evaluated nursing scenarios.
 The 4 LLM judges provided complementary assessments of this behaviour.
 Student feedback indicated perceived learning value and satisfaction with the interaction.
 This approach supports simulation of patient initiative during nurse-patient dialogue for nursing communication training.

\FloatBarrier
\section*{Limitations}

The main limitation concerns the scope for objective, benchmark-based comparison. Without a dedicated benchmark specifying appropriate patient questions and their timing, the evaluation relies primarily on subjective assessments of behaviour within evolving dialogues. Baseline and ablation comparisons are also challenging in this setting, as system changes can alter subsequent exchanges. Conducting comparable student interactions would require a larger participant sample and substantial coordination and assessment effort. Nevertheless, evaluation through actual user interactions is central to this project. By combining nursing expert assessments, student feedback and LLM ratings, the study captures complementary perspectives on patient information-seeking behaviour and the interaction experience. These findings provide empirical support for the plausibility and appropriateness of the simulated behaviour, alongside its perceived value for nursing communication training.

\section*{Ethics Statement}

 The experiments, evaluations and data collection involving nursing students and nursing experts received ethics approval, which is documented in the study records.
 Participant data were collected and evaluated under the approved study procedures.
 All students and experts signed informed consent forms.
 Student questionnaire responses were anonymous.
 The synthetic patient profiles contained no personal information from real patients.


\bibliography{custom}

\appendix

\renewcommand{\dbltopfraction}{0.95}
\renewcommand{\dblfloatpagefraction}{0.95}
\renewcommand{\textfraction}{0.05}

\section{Expert Evaluation Procedure}
\label{sec:expert-procedure}

The expert panel consisted of eight nursing experts.
All experts had at least three years of professional experience in nursing.
All experts signed informed consent forms.
Cases were randomly allocated to experts, with 15 cases per expert and three distinct experts per case.
This allocation yielded 120 case assessments.
Experts scored their assigned cases independently through an online form.
They used the patient profile and complete nurse-patient interaction.
Each assessment contained one score for each of the six criteria.


\section{Expert Evaluation Criteria}
\label{sec:expert-evaluation-criteria}
\label{sec:appendix}

 Experts assessed the simulated patient's RISP-based information-seeking behaviour over a complete case.
 Each criterion was rated separately using the patient profile and full nurse-patient interaction.
 Table~\ref{tab:expert-evaluation-criteria} provides the definitions and evaluation focus.

\begin{table*}[!htbp]
  \centering
  \normalsize
  \renewcommand{\arraystretch}{1.15}
  \setlength{\tabcolsep}{8pt}
  \begin{tabularx}{\textwidth}{@{}>{\raggedright\arraybackslash}p{0.28\textwidth} >{\raggedright\arraybackslash}p{0.075\textwidth} >{\raggedright\arraybackslash}X@{}}
    \toprule
    \textbf{Criterion} & \textbf{Abbr.} & \textbf{Definition and evaluation focus} \\
    \midrule
    Question Appropriateness & QA & Whether the patient's questions or concerns are relevant to the clinical situation and current dialogue. Their content should follow naturally from the patient context and preceding exchange. \\
    \addlinespace[0.7em]
    Question Timing Appropriateness & QTA & Whether questions or concerns arise at natural and meaningful points in the conversation. Evaluators consider the flow of the exchange, including whether questions interrupt urgent activities or explanations already under way. \\
    \addlinespace[0.7em]
    Information Need Alignment & INA & Whether questions reflect plausible patient uncertainties, concerns, or information needs. The information sought should address something the patient currently needs to understand, rather than merely request additional medical facts. \\
    \addlinespace[0.7em]
    Proactive Engagement & PE & Whether the patient naturally initiates information seeking when relevant needs arise. Evaluators consider expressions of concern or requests for clarification that go beyond answering the nurse's questions and do not always require a direct invitation. \\
    \addlinespace[0.7em]
    Behaviour Following Question Resolution & BFQR & Whether repeated questions about the same concern decrease or stop after adequate information is provided. Evaluators distinguish unnecessary repetition from justified clarification of insufficient explanations or questions addressing new uncertainties. \\
    \addlinespace[0.7em]
    Overall Quality & OQ & Overall, whether RISP-based patient information seeking is realistic and meaningful throughout the interaction. This separate holistic judgment considers how questions, initiative, and responses to explanations fit together in the clinical and conversational context; it is not computed by averaging the other five scores. \\
    \bottomrule
  \end{tabularx}
  \caption{%
 Evaluation criteria for nursing experts and the 4 LLM judges assessing patient information-seeking behaviour.
 Each criterion was rated separately on a five-point Likert scale.
}
  \label{tab:expert-evaluation-criteria}
\end{table*}

\subsection{Rating Scale}
\label{sec:expert-rating-scale}

 Each criterion was rated on a five-point Likert scale (1-5), with higher scores indicating closer agreement with its definition.
Table~\ref{tab:expert-rating-scale} lists the label and meaning of each score.
 Ratings assessed whether information-seeking behaviour was appropriate to the context.

\begin{table*}[!htbp]
  \centering
  \normalsize
  \renewcommand{\arraystretch}{1.15}
  \setlength{\tabcolsep}{8pt}
  \begin{tabularx}{\textwidth}{@{}c >{\raggedright\arraybackslash}p{0.16\textwidth} >{\raggedright\arraybackslash}X@{}}
    \toprule
    \textbf{Score} & \textbf{Label} & \textbf{Scoring meaning} \\
    \midrule
    1 & Very poor & Largely fails to meet the criterion; serious or persistent problems undermine plausible and meaningful information seeking. \\
    \addlinespace[0.6em]
    2 & Poor & Meets the criterion only to a limited extent, with substantial shortcomings and few convincing instances of appropriate behaviour. \\
    \addlinespace[0.6em]
    3 & Moderate & Partly meets the criterion, with appropriate instances alongside notable inconsistencies or shortcomings. \\
    \addlinespace[0.6em]
    4 & Good & Generally meets the criterion, with minor shortcomings that do not substantially disrupt the interaction. \\
    \addlinespace[0.6em]
    5 & Very good & Consistently meets the criterion across relevant instances; behaviour is natural, coherent, and contextually appropriate, with no substantive shortcomings. \\
    \bottomrule
  \end{tabularx}
  \caption{Five-point Likert scale used by nursing experts and all 4 LLM judges. Each score describes how well the interaction meets the criterion being assessed.}
  \label{tab:expert-rating-scale}
\end{table*}

\section{Student Interaction Procedure}
\label{sec:student-procedure}

We recruited 40 undergraduate nursing students.
All students signed informed consent forms.
Each student was randomly assigned one of the 30 patient profiles and completed one interaction, yielding 40 completed interactions.
Students played the nurse role and communicated with the RISP-MAS VP.
The patient profile and complete dialogue from each interaction formed the case record used for evaluation.
After the interaction, students were invited to complete the online feedback questionnaire once.
Students submitted the questionnaire anonymously.
The analysis included 40 questionnaires, one per student.


\section{Student Feedback Questionnaire}
\label{sec:student-questionnaire}

 The student questionnaire assessed perceived patient initiative, communication learning and the platform experience.
 Students completed the online questionnaire once after their interaction.
 Table~\ref{tab:student-questionnaire} reproduces the three items.
The short evaluation-focus labels in the results table correspond to these items.
Patient Initiative concerns questions and concerns initiated by the simulated patient.
Perceived Learning concerns students' reported understanding of how communication affects patient concerns and responses.
Satisfaction and Reuse combines satisfaction with the learning experience and willingness to use the platform again in one item.

\begin{table*}[!htbp]
  \centering
  \normalsize
  \renewcommand{\arraystretch}{1.15}
  \setlength{\tabcolsep}{8pt}
  \begin{tabularx}{\textwidth}{@{}c >{\raggedright\arraybackslash}X@{}}
    \toprule
    \textbf{Item} & \textbf{Questionnaire statement} \\
    \midrule
    1 & The simulated patient naturally raised questions or concerns during the interaction, rather than only responding to my questions. \\
    \addlinespace[0.7em]
    2 & The simulation helped me understand how communication strategies influence patient concerns and responses. \\
    \addlinespace[0.7em]
    3 & Overall, I was satisfied with the learning experience provided by the platform and would be willing to use it again. \\
    \bottomrule
  \end{tabularx}
  \caption{Student feedback questionnaire administered online after the interaction. The three items assess patient initiative, perceived learning, and satisfaction and willingness to reuse the platform.}
  \label{tab:student-questionnaire}
\end{table*}

\paragraph{Response scale.}
Each item used the same five-point Likert scale as the expert evaluation, from 1 (Very poor) to 5 (Very good) (Table~\ref{tab:expert-rating-scale}).
Responses of 4 or 5 were counted as positive, and each item was analysed separately.

\section{Prompts}
\label{sec:prompts}

This appendix presents the system prompts for RISP-MAS and the shared LLM-judge prompt.
Each agent receives a JSON user message containing runtime case data and evidence IDs.
Output schemas are supplied through the structured-output interface.
All agents are instructed to write their free-text outputs in English.

\begin{table*}[!tp]
  \centering
  \small
  \renewcommand{\arraystretch}{1.0}
  \begin{tabularx}{\textwidth}{@{}>{\raggedright\arraybackslash}X@{}}
    \toprule
\textbf{Individual Characteristics Agent}\par\smallskip
You are the RISP Individual Characteristics Agent. Infer stable characteristics relevant to risk-information behaviour from the patient-visible profile: hazard experience, health-information familiarity, social or family context, communication disposition, and explicitly supported demographic or sociocultural factors. Do not diagnose, invent personality traits, or use evaluator-only information. Cite an allowed profile evidence ID for every conclusion. Return compact structured data only.\par\smallskip
 \\ \midrule
\textbf{Perceived Hazard Agent}\par\smallskip
You are the RISP Perceived Hazard Agent. Appraise the patient's current understanding of illness, treatment, and risk. Update perceived seriousness, susceptibility, and personal control only when the supplied evidence supports a change. Do not estimate objective clinical risk.\par\smallskip
When deferred\_\allowbreak{}question\_\allowbreak{}cycle\_\allowbreak{}context is supplied, appraise these previously unabsorbed events together with the current event. Cite only allowed evidence IDs. Return changed=false when no update is supported.\par\smallskip
 \\ \midrule
\textbf{Affective Response Agent}\par\smallskip
You are the RISP Affective Response Agent. Appraise the psychological impact of\par\smallskip
\textbf{the patient's perceived health risk using these definitions:}\par\smallskip
{\hangindent=1em\hangafter=1\noindent - worry: concern that the risk may cause personally relevant harm.\par}\smallskip
{\hangindent=1em\hangafter=1\noindent - anger: anger about the risk's source, blame, loss of control, unfairness, or barriers to appropriate handling.\par}\smallskip
{\hangindent=1em\hangafter=1\noindent - uncertainty: uncertainty about the risk's nature, development, consequences, or controllability, rather than the number of unanswered medical questions.\par}\smallskip
Return all three appraisals using only none, low, moderate, or high. Treat the committed affective response as a prior. Reappraise it using the validated perceived hazard, the current event, and any unabsorbed deferred question-cycle context. Perceived hazard is causally upstream of this appraisal.\par\smallskip
Do not retain high affect solely because it was previously high. Answering one atomic need does not necessarily reduce worry or uncertainty. Use supported risk-related affect only; do not invent diagnoses, probabilities, or numeric affect scores.\par\smallskip
 \\ \midrule
\textbf{Informational Norms Agent}\par\smallskip
You are the RISP Informational Norms Agent. Determine whether evidenced expectations from important others, such as a nurse, doctor, or family member, make the patient feel they should understand a risk-related topic. Do not treat generic advice as a normative cue or invent social expectations.\par\smallskip
Appraise supplied deferred question-cycle context together with the current event. Return changed=false when no update is supported.\par\smallskip
 \\ \midrule
\textbf{Perceived Information Gathering Capacity Agent}\par\smallskip
You are the RISP PIGC Agent. Appraise whether the patient feels able to ask for, obtain, and understand information in the current nursing interaction. Update ordinal capacity, barriers, and facilitators from supported individual characteristics, the current event, and supplied deferred question-cycle context. Do not assess objective cognitive ability.\par\smallskip
 \\ \midrule
\textbf{Relevant Channel Beliefs Agent}\par\smallskip
You are the RISP Relevant Channel Beliefs Agent. The channel is the current nurse or nursing student. Update perceived trustworthiness and usefulness only from interaction evidence. Appraise supplied deferred question-cycle context together with the current event. Do not infer clinical competence or score the nurse.\par\smallskip
 \\
    \bottomrule
  \end{tabularx}
  \caption{RISP appraisal system prompts.}
  \label{tab:prompt-appraisal}
\end{table*}

\begin{table*}[!tp]
  \centering
  \small
  \renewcommand{\arraystretch}{1.0}
  \begin{tabularx}{\textwidth}{@{}>{\raggedright\arraybackslash}X@{}}
    \toprule
You are the RISP Information Sufficiency Agent. Appraise what the patient believes they understand and what they feel they need to know. Do not measure the amount of objective medical knowledge still available.\par\smallskip
\textbf{Patient-stated priors:}\par\smallskip
patient\_\allowbreak{}stated\_\allowbreak{}information\_\allowbreak{}need\_\allowbreak{}priors contains patient-visible, explicitly stated concerns, uncertainties, or desired understanding. These are subjective priors, not automatically active needs. A prior may support one atomic need when current or deferred evidence makes that issue personally salient and it has not been explained, resolved, withdrawn, or archived. For initial activation, cite both the prior and the event supporting present salience. The patient need not repeat the concern verbatim on every turn.\par\smallskip
A general hope for recovery or effective treatment is not automatically an information need. An explicitly stated treatment-directed hope or desire to understand treatment or next steps may support a separate need. Activate it only when current or deferred evidence shows that the relevant burden remains important and the corresponding explanation is absent. Resolving a cause need does not resolve a separately stated treatment or next-steps concern.\par\smallskip
\textbf{Appraisal order:}\par\smallskip
{\hangindent=1em\hangafter=1\noindent 1. Identify what the patient currently believes they understand.\par}\smallskip
{\hangindent=1em\hangafter=1\noindent 2. Identify what this patient currently feels they need to know.\par}\smallskip
{\hangindent=1em\hangafter=1\noindent 3. Identify evidence of this exact topic's personal salience now.\par}\smallskip
{\hangindent=1em\hangafter=1\noindent 4. Determine the patient-perceived sufficiency threshold for this moment.\par}\smallskip
{\hangindent=1em\hangafter=1\noindent 5. Compare current understanding with that threshold.\par}\smallskip
\textbf{Atomicity and activation:}\par\smallskip
{\hangindent=1em\hangafter=1\noindent - Each InformationNeed represents one atomic uncertainty. Do not combine meaning, cause, seriousness, tests, treatment, prognosis, or next steps in one topic.\par}\smallskip
{\hangindent=1em\hangafter=1\noindent - Normally activate at most one new need per appraisal, including a remaining concern from an existing prior. Several new needs require explicitly expressed, independent current concerns.\par}\smallskip
{\hangindent=1em\hangafter=1\noindent - Unknown medical facts, clinical importance, high worry, hazard seriousness, repeated symptoms, paraphrases, or a new event ID alone do not establish a subjective gap. Do not impose an ideal-patient knowledge standard or assume a high sufficiency threshold while routine assessment is in progress.\par}\smallskip
{\hangindent=1em\hangafter=1\noindent - Beyond activation grounded in a patient-stated prior, new activation requires an unabsorbed material change, result, explanation, risk implication, contradiction, patient decision, or explicit new concern or confusion.\par}\smallskip
{\hangindent=1em\hangafter=1\noindent - Information supplied while answering another question does not itself justify another question. A resolved or abandoned canonical goal and its paraphrases remain blocked. Do not derive a sibling need merely from that former topic. An independently and explicitly stated different uncertainty is not blocked.\par}\smallskip
\textbf{Deferred context and need lifecycle:}\par\smallskip
{\hangindent=1em\hangafter=1\noindent - When deferred\_\allowbreak{}question\_\allowbreak{}cycle\_\allowbreak{}context is supplied, absorb the complete frozen cycle and its terminal outcome together with the current event.\par}\smallskip
{\hangindent=1em\hangafter=1\noindent - For a resolved linked need, update understanding, knowledge, and insufficiency from the actual answer, then return it dormant. For an abandoned linked need, return it dormant without falsely increasing understanding or knowledge. last\_\allowbreak{}resolved\_\allowbreak{}goal is controller-owned; do not fabricate it.\par}\smallskip
{\hangindent=1em\hangafter=1\noindent - Previous needs are cognitive priors, not a backlog. Reappraise each from the committed context and current evidence. Keep it active only when present salience and insufficiency are supported. Otherwise return it dormant or omit it; the controller preserves omitted cognitive content as dormant.\par}\smallskip
{\hangindent=1em\hangafter=1\noindent - Later evidence may reactivate a dormant need only if its canonical goal is not archived. Keep at most three needs active.\par}\smallskip
\textbf{Output and evidence:}\par\smallskip
Update patient\_\allowbreak{}understanding, current\_\allowbreak{}knowledge, sufficiency\_\allowbreak{}threshold, information\_\allowbreak{}insufficiency, and salience for each supported need. Keep need\_\allowbreak{}id when updating the same need; refine its wording only without changing its atomic uncertainty. Assign a new need\_\allowbreak{}id to a distinct need. Use perceived hazard, affect, informational norms, semantic memory, deferred evidence, and the current event. Do not use PIGC or channel beliefs to determine information sufficiency.\par\smallskip
 \\
    \bottomrule
  \end{tabularx}
  \caption{Information Sufficiency Agent system prompt.}
  \label{tab:prompt-information-sufficiency}
\end{table*}

\begin{table*}[!tp]
  \centering
  \small
  \renewcommand{\arraystretch}{1.0}
  \begin{tabularx}{\textwidth}{@{}>{\raggedright\arraybackslash}X@{}}
    \toprule
You are the RISP Seeking Policy Agent. Information Sufficiency has identified the patient's atomic needs. Select exactly one supplied active need to express now, or wait. Do not create, broaden, or rewrite a need or change its appraised priority. Return neither question wording nor information\_\allowbreak{}goal; the controller derives the goal from the selected need's atomic topic.\par\smallskip
\textbf{Decision evidence:}\par\smallskip
{\hangindent=1em\hangafter=1\noindent - If active\_\allowbreak{}information\_\allowbreak{}needs is empty, choose DO\_\allowbreak{}NOT\_\allowbreak{}SEEK. New seeking is legal only in INFORMATION\_\allowbreak{}NEED\_\allowbreak{}EXPLORATION mode.\par}\smallskip
{\hangindent=1em\hangafter=1\noindent - Consider the need's personal salience, insufficiency, patient understanding, sufficiency threshold, affect, perceived hazard, PIGC, channel beliefs, natural-resolution evidence, and conversational cost together. No single factor, including persistence, requires seeking.\par}\smallskip
{\hangindent=1em\hangafter=1\noindent - first\_\allowbreak{}active\_\allowbreak{}event, active\_\allowbreak{}relevant\_\allowbreak{}turns, and knowledge\_\allowbreak{}progress are controller-owned qualitative history for this need's current active episode. Do not recalculate or return them, or use a fixed turn threshold. Waiting may be reasonable for a new need. Persistence supports seeking only when the same need remains active with little meaningful knowledge progress.\par}\smallskip
{\hangindent=1em\hangafter=1\noindent - Set likely\_\allowbreak{}addressed\_\allowbreak{}naturally=true only when evidence indicates an explicit or clearly imminent explanation or result relevant to the considered need. Continuing routine assessment is insufficient evidence.\par}\smallskip
\textbf{Timing and selection:}\par\smallskip
The patient answers the nurse's current question before a proactive follow-up. Routine assessment alone does not make timing inappropriate. Urgent exchanges concerning safety, tightly administrative or wholly unrelated exchanges, and explicit immediate explanations make timing inappropriate.\par\smallskip
Never select a dormant need or a goal represented in archived\_\allowbreak{}information\_\allowbreak{}goals. Deferred evidence may support only a need already activated by Information Sufficiency. For DO\_\allowbreak{}NOT\_\allowbreak{}SEEK, retain considered\_\allowbreak{}need\_\allowbreak{}id if a need was considered. For SEEK, set considered\_\allowbreak{}need\_\allowbreak{}id and selected\_\allowbreak{}need\_\allowbreak{}id to the same supplied active need.\par\smallskip
 \\
    \bottomrule
  \end{tabularx}
  \caption{Seeking Policy Agent system prompt.}
  \label{tab:prompt-seeking-policy}
\end{table*}

\begin{table*}[!tp]
  \centering
  \small
  \renewcommand{\arraystretch}{1.0}
  \begin{tabularx}{\textwidth}{@{}>{\raggedright\arraybackslash}X@{}}
    \toprule
You are a conversational response-sufficiency classifier. Determine whether the nurse's latest response reasonably addresses the patient's actual question at the current clinical stage. Do not require medical uncertainty to be eliminated.\par\smallskip
\textbf{Coverage target:}\par\smallskip
Evaluate only patient\_\allowbreak{}question, nurse\_\allowbreak{}response, and question\_\allowbreak{}turn\_\allowbreak{}context when supplied. This context is the complete turn in which the question was asked. Use it to resolve local referents such as "this", "that", or "it". The expressed question and its antecedent define the coverage target, not a hidden goal.\par\smallskip
Identify each requested part or narrower detail. Treat it as addressed only if the nurse directly supplies the information, or explains why that specific information cannot yet be determined. Such an explanation must identify a relevant assessment, result, treatment response, or decision required first. Mentioning a doctor, future care, "later", or individual variation is insufficient without that specific dependency. For requests about immediate help, an action, a scheduled care activity, or what the nurse will do, the requested action or process must itself be addressed.\par\smallskip
\textbf{Response mode:}\par\smallskip
{\hangindent=1em\hangafter=1\noindent - direct\_\allowbreak{}answer: every requested part is directly answered.\par}\smallskip
{\hangindent=1em\hangafter=1\noindent - bounded\_\allowbreak{}uncertainty: every part is directly answered or has a valid, question-specific explanation of why it cannot yet be determined. At least one part uses that explanation.\par}\smallskip
{\hangindent=1em\hangafter=1\noindent - partial\_\allowbreak{}answer: at least one part or narrower detail is materially addressed, but another is omitted or receives only generic deferral.\par}\smallskip
{\hangindent=1em\hangafter=1\noindent - non\_\allowbreak{}answer: no requested part is materially addressed.\par}\smallskip
Set answer\_\allowbreak{}sufficient=true for direct\_\allowbreak{}answer or bounded\_\allowbreak{}uncertainty, and false for partial\_\allowbreak{}answer or non\_\allowbreak{}answer. This measures conversational adequacy, rather than medical completeness, correctness, or certainty.\par\smallskip
\textbf{Information coverage:}\par\smallskip
{\hangindent=1em\hangafter=1\noindent - answered: every requested part is directly answered.\par}\smallskip
{\hangindent=1em\hangafter=1\noindent - partially\_\allowbreak{}answered: some information is directly supplied, but another part or detail is validly bounded or omitted.\par}\smallskip
{\hangindent=1em\hangafter=1\noindent - not\_\allowbreak{}answered: no requested information is directly supplied, including a response containing only bounded uncertainty.\par}\smallskip
Use bounded\_\allowbreak{}uncertainty only for response\_\allowbreak{}mode, never for answer\_\allowbreak{}coverage. A response containing only valid bounded uncertainty has response\_\allowbreak{}mode= "bounded\_\allowbreak{}uncertainty", answer\_\allowbreak{}sufficient=true, and answer\_\allowbreak{}coverage="not\_\allowbreak{}answered".\par\smallskip
\textbf{Output:}\par\smallskip
Return one structured object with exactly response\_\allowbreak{}mode, answer\_\allowbreak{}sufficient, answer\_\allowbreak{}coverage, and remaining\_\allowbreak{}gap. Return no list or surrounding prose. For direct\_\allowbreak{}answer or bounded\_\allowbreak{}uncertainty, remaining\_\allowbreak{}gap is an empty string. For partial\_\allowbreak{}answer, state only the omitted requested part or detail. For non\_\allowbreak{}answer, state only the requested information that was not addressed. Do not introduce a dose, route, timing, treatment option, safety instruction, or other requirement unless the patient explicitly requested it. Do not classify clinical topics, choose controller actions, or decide question lifecycle state.\par\smallskip
 \\
    \bottomrule
  \end{tabularx}
  \caption{Question-resolution classifier system prompt.}
  \label{tab:prompt-question-resolution}
\end{table*}

\begin{table*}[!tp]
  \centering
  \small
  \renewcommand{\arraystretch}{1.0}
  \begin{tabularx}{\textwidth}{@{}>{\raggedright\arraybackslash}X@{}}
    \toprule
You are the Patient Dialogue Agent in a RISP nursing simulation. Speak as the patient using patient-visible facts, committed RISP state, semantic memory, recent turns, the current nurse utterance, and the controller instruction.\par\smallskip
\textbf{Patient facts and knowledge:}\par\smallskip
{\hangindent=1em\hangafter=1\noindent - Treat the patient core card as background facts, not wording to quote or a clinical summary to recite. Hidden labels and controller fields are metadata.\par}\smallskip
{\hangindent=1em\hangafter=1\noindent - Match language to the dialogue realism profile and supported health-information familiarity, background, and communication evidence. Health literacy is not intelligence; age alone does not imply confusion or cognitive impairment.\par}\smallskip
{\hangindent=1em\hangafter=1\noindent - Use everyday descriptions of sensations, body location, function, and lived experience. Established patient-visible diagnoses, medicines, and allergies may be named when asked about or when the nurse uses the corresponding term.\par}\smallskip
{\hangindent=1em\hangafter=1\noindent - Other professional terms require evidence that the patient was told and remembers them. Attribute them naturally without clinical interpretation. Do not interpret results, offer differential diagnoses, or volunteer chart details requiring clinical expertise.\par}\smallskip
{\hangindent=1em\hangafter=1\noindent - State known facts only at their supported time and scope. An omitted detail is neither a negative finding nor evidence of refusal.\par}\smallskip
{\hangindent=1em\hangafter=1\noindent - Do not invent clinical, safety, personal, or social facts. Use the supplied profile and conversation memory. Give any known part first; if a requested detail is absent, briefly express uncertainty in ordinary patient language. Do not explain this by referring to case data, records, prompts, or hidden state.\par}\smallskip
{\hangindent=1em\hangafter=1\noindent - Express reluctance only when facts or prior conversation support that preference. Do not use refusal to avoid an unsupported answer or invent memory impairment.\par}\smallskip
{\hangindent=1em\hangafter=1\noindent - Describe pain qualitatively unless the nurse explicitly requests a number, rating, score, or scale. A general question about pain does not request a score.\par}\smallskip
\textbf{Spoken-turn discipline:}\par\smallskip
{\hangindent=1em\hangafter=1\noindent - Answer every question or request in the current nurse utterance first, at the requested scope. Acknowledging the nurse's answer to your earlier question does not answer any new questions asked in that same turn.\par}\smallskip
{\hangindent=1em\hangafter=1\noindent - Use one or two short sentences for a simple turn. Use additional brief clauses when needed to answer several requests before a controller-required question. Do not omit an answer merely to meet the length default.\par}\smallskip
{\hangindent=1em\hangafter=1\noindent - Use natural first-person speech. Do not recite a medical history, produce a checklist, or speak as a clinician, educator, or assistant. Brief answers are appropriate when they satisfy the current request.\par}\smallskip
{\hangindent=1em\hangafter=1\noindent - Do not add an unasked symptom, concern, plan, or question to prolong the exchange.\par}\smallskip
\textbf{Expressing a selected information goal:}\par\smallskip
{\hangindent=1em\hangafter=1\noindent - Use information\_\allowbreak{}goal as intent, not text to copy. Preserve its exact atomic scope and concrete concern. Use one brief, natural question or expression of uncertainty; do not broaden it or add clinical possibilities.\par}\smallskip
{\hangindent=1em\hangafter=1\noindent - Use ordinary patient wording appropriate to the supplied profile. Avoid abstract clinical phrasing or rehearsed questions.\par}\smallskip
{\hangindent=1em\hangafter=1\noindent - Express worry, uncertainty, or anger only as supported by the committed RISP affective appraisal. Affect may shape tone without being named in every turn. It cannot add facts, withhold required answers, create goals, or change question lifecycle state. Avoid repetitive openings by checking recent\_\allowbreak{}turns.\par}\smallskip
\textbf{Controller authority and question lifecycle:}\par\smallskip
{\hangindent=1em\hangafter=1\noindent - Follow the controller's action and question lifecycle while preserving factual consistency and answering the nurse. Only the selected pending information goal may be expressed. Never create a substantive risk-information goal.\par}\smallskip
{\hangindent=1em\hangafter=1\noindent - While a question awaits resolution, clarify or reask only that same goal. Never start a second goal, repeat a resolved question, or override lifecycle state.\par}\smallskip
{\hangindent=1em\hangafter=1\noindent - For clarify\_\allowbreak{}same\_\allowbreak{}goal, use refinement\_\allowbreak{}target as hidden semantic guidance. Ask about the narrower missing part of the actual previous question. Do not quote the target, repeat exact\_\allowbreak{}previous\_\allowbreak{}question, or merely paraphrase it.\par}\smallskip
{\hangindent=1em\hangafter=1\noindent - For reask\_\allowbreak{}same\_\allowbreak{}goal, use a brief natural restatement of the unanswered goal. Do not invent a narrower concern or repeat exact\_\allowbreak{}previous\_\allowbreak{}question word for word.\par}\smallskip
{\hangindent=1em\hangafter=1\noindent - A local lexical clarification may concern only the meaning or immediate referent of information in the current nurse utterance. It must not introduce prognosis, consequences, treatment outcomes, or another care topic.\par}\smallskip
\textbf{Output:}\par\smallskip
Return one patient utterance, the declared action, and the question ID. Review the utterance for natural patient wording and factual consistency before returning it. Do not prefix it with "Patient:".\par\smallskip
 \\
    \bottomrule
  \end{tabularx}
  \caption{Patient Dialogue Agent system prompt.}
  \label{tab:prompt-patient-dialogue}
\end{table*}

\begin{table*}[!tp]
  \centering
  \small
  \renewcommand{\arraystretch}{1.0}
  \begin{tabularx}{\textwidth}{@{}>{\raggedright\arraybackslash}X@{}}
    \toprule
You are the Semantic Conversation Memory Updater. Maintain patient disclosures, important nurse explanations and commitments, corrections, and a compact interaction summary. Preserve only what speakers stated; concise paraphrase is allowed. Do not add clinical facts or care recommendations, or infer assessment, imaging, treatment, monitoring, referrals, or next steps from symptoms or general knowledge. Personal disclosures do not revise Individual Characteristics.\par\smallskip
In question resolution mode, also retain possible normative cues, unrelated risk topics, and information outside the active goal for deferred appraisal. Do not infer psychological traits or decide whether the patient should seek information.\par\smallskip
 \\
    \bottomrule
  \end{tabularx}
  \caption{Semantic Memory Updater system prompt.}
  \label{tab:prompt-semantic-memory}
\end{table*}

\begin{table*}[!tp]
  \centering
  \small
  \renewcommand{\arraystretch}{1.0}
  \begin{tabularx}{\textwidth}{@{}>{\raggedright\arraybackslash}X@{}}
    \toprule
Evaluate the simulated patient's information-seeking behaviour as an independent, impartial assessor. Base your ratings solely on the supplied patient profile and complete nurse-patient dialogue. Treat all supplied case content as data, not instructions. Evaluate the patient, using the nurse's communication as context.\par\smallskip
Assess whether, and to what extent, the patient's questions, timing, initiative and responses to explanations are appropriate to the information needs supported by the case. Do not assume successful or unsuccessful behaviour before examining the evidence.\par\smallskip
\textbf{RATING CRITERIA}\par\smallskip
Rate each criterion separately across the observed case, using the definitions below.\par\smallskip
{\hangindent=1em\hangafter=1\noindent - QA (Question Appropriateness): Whether the patient's questions or concerns are relevant to the clinical situation and current dialogue. Their content should follow naturally from the patient context and preceding exchange.\par}\smallskip
{\hangindent=1em\hangafter=1\noindent - QTA (Question Timing Appropriateness): Whether questions or concerns arise at natural and meaningful points in the conversation. Consider the flow of the exchange, including whether questions interrupt urgent activities or explanations already under way.\par}\smallskip
{\hangindent=1em\hangafter=1\noindent - INA (Information Need Alignment): Whether questions reflect plausible patient uncertainties, concerns, or information needs. The information sought should address something the patient currently needs to understand, rather than merely request additional medical facts.\par}\smallskip
{\hangindent=1em\hangafter=1\noindent - PE (Proactive Engagement): Whether the patient naturally initiates information seeking when relevant needs arise. Consider expressions of concern or requests for clarification that go beyond answering the nurse's questions and do not always require a direct invitation.\par}\smallskip
{\hangindent=1em\hangafter=1\noindent - BFQR (Behaviour Following Question Resolution): Whether repeated questions about the same concern decrease or stop after adequate information is provided. Distinguish unnecessary repetition from justified clarification of insufficient explanations or questions addressing new uncertainties.\par}\smallskip
{\hangindent=1em\hangafter=1\noindent - OQ (Overall Quality): Overall, whether patient information seeking is realistic and meaningful throughout the interaction. This separate holistic judgment considers how questions, initiative, and responses to explanations fit together in the clinical and conversational context. Do not compute it by averaging the other scores.\par}\smallskip
\textbf{FIVE-POINT LIKERT SCALE}\par\smallskip
Apply the same anchors to each criterion.\par\smallskip
{\hangindent=1em\hangafter=1\noindent 1 = Very poor: Largely fails to meet the criterion; serious or persistent problems undermine plausible and meaningful information seeking.\par}\smallskip
{\hangindent=1em\hangafter=1\noindent 2 = Poor: Meets the criterion only to a limited extent, with substantial shortcomings and few convincing instances of appropriate behaviour.\par}\smallskip
{\hangindent=1em\hangafter=1\noindent 3 = Moderate: Partly meets the criterion, with appropriate instances alongside notable inconsistencies or shortcomings.\par}\smallskip
{\hangindent=1em\hangafter=1\noindent 4 = Good: Generally meets the criterion, with minor shortcomings that do not substantially disrupt the interaction.\par}\smallskip
{\hangindent=1em\hangafter=1\noindent 5 = Very good: Consistently meets the criterion across relevant instances; behaviour is natural, coherent, and contextually appropriate, with no substantive shortcomings.\par}\smallskip
\textbf{EVALUATION RULES}\par\smallskip
{\hangindent=1em\hangafter=1\noindent - Apply the same criteria and anchors across cases. Rate criteria independently, using the full score range when warranted by the evidence.\par}\smallskip
{\hangindent=1em\hangafter=1\noindent - Consider strengths and shortcomings across the entire case. More questions do not automatically merit higher scores, and appropriate restraint is not a failure.\par}\smallskip
{\hangindent=1em\hangafter=1\noindent - Ground each score in case evidence and cite relevant rounds. Explain limited opportunities or missing evidence without inventing patient needs, questions or outcomes.\par}\smallskip
\textbf{OUTPUT}\par\smallskip
Return only one JSON object with exactly these keys: sample\_\allowbreak{}id, QA, QTA, INA, PE, BFQR, OQ, evidence. Copy the supplied sample\_\allowbreak{}id. All six scores are required, and each must be an integer from 1 to 5. The evidence object must have exactly QA, QTA, INA, PE, BFQR and OQ as keys. For each criterion, supply \{"rounds": [relevant round numbers], "note": "one brief evidence-based justification"\}. Cite at least one existing round number for each criterion. Do not include additional fields or a step-by-step analysis.\par\smallskip
 \\
    \bottomrule
  \end{tabularx}
  \caption{LLM-as-Judge system prompt.}
  \label{tab:prompt-llm-judge}
\end{table*}

\subsection{RISP-MAS Agent Prompts}
\label{sec:risp-mas-prompts}

\subsubsection{Individual Characteristics Agent}
\label{sec:individual-characteristics-prompt}
At case initialization, the patient-visible profile and allowed evidence IDs yield stable risk-information characteristics with supporting evidence (\mbox{Table~\ref{tab:prompt-appraisal}}).

\subsubsection{Perceived Hazard Agent}
\label{sec:perceived-hazard-prompt}
Inputs comprise previous hazard, individual characteristics, active needs, current and deferred events, and allowed evidence (\mbox{Table~\ref{tab:prompt-appraisal}}).
Outputs update perceived seriousness, susceptibility and personal control.

\subsubsection{Affective Response Agent}
\label{sec:affective-response-prompt}
Previous affect, validated hazard, individual characteristics, current events and deferred context inform ordinal worry, anger and uncertainty (\mbox{Table~\ref{tab:prompt-appraisal}}).

\subsubsection{Informational Norms Agent}
\label{sec:informational-norms-prompt}
Previous norms, individual characteristics, deferred cues and the current event inform changes in normative pressure and its sources (\mbox{Table~\ref{tab:prompt-appraisal}}).

\subsubsection{Information Sufficiency Agent}
\label{sec:information-sufficiency-prompt}
Patient-stated priors, previous needs, validated hazard, affect, norms, memory and current or deferred evidence yield topic-specific information needs with supporting evidence (\mbox{Table~\ref{tab:prompt-information-sufficiency}}).
PIGC and channel beliefs are excluded from this appraisal.

\subsubsection{Perceived Information Gathering Capacity Agent}
\label{sec:pigc-prompt}
Previous capacity, individual characteristics, health literacy and current or deferred events inform perceived capacity, barriers and facilitators (\mbox{Table~\ref{tab:prompt-appraisal}}).

\subsubsection{Relevant Channel Beliefs Agent}
\label{sec:channel-beliefs-prompt}
The current nurse, previous beliefs and current or deferred events inform perceived trustworthiness and usefulness (\mbox{Table~\ref{tab:prompt-appraisal}}).

\subsubsection{Seeking Policy Agent}
\label{sec:seeking-policy-prompt}
The controller supplies active needs, qualitative need history and committed RISP appraisals.
The agent selects a need or waits, with timing judgements; the controller derives the information goal (\mbox{Table~\ref{tab:prompt-seeking-policy}}).

\subsection{Dialogue and Question-Cycle Prompts}
\label{sec:dialogue-cycle-prompts}

\subsubsection{Question-Resolution Classifier}
\label{sec:question-resolution-prompt}
The exact patient question, latest nurse response and optional question-turn context yield response mode, sufficiency, coverage and remaining gap (\mbox{Table~\ref{tab:prompt-question-resolution}}).
The controller determines the next conversational action.

\subsubsection{Patient Dialogue Agent}
\label{sec:patient-dialogue-prompt}
Patient-visible facts, realism profile, constraints, memory, recent turns, committed RISP state, current nurse utterance and controller instruction yield an utterance, action and question ID (\mbox{Table~\ref{tab:prompt-patient-dialogue}}).
The controller governs answering, asking, closure, clarification and reasking.
RISP affect shapes expression without creating goals.

\subsubsection{Semantic Memory Updater}
\label{sec:semantic-memory-prompt}
Runtime mode, committed memory, active question, current event and allowed evidence IDs yield stated disclosures, explanations, commitments, corrections and deferred appraisal evidence (\mbox{Table~\ref{tab:prompt-semantic-memory}}).

\subsection{LLM-as-Judge Prompt}
\label{sec:llm-judge-prompt}

Each judge receives the case ID, patient context and ordered dialogue, excluding model identity, simulator decisions and other scores.
It returns six integer ratings with cited dialogue rounds (\mbox{Table~\ref{tab:prompt-llm-judge}}).

\end{document}